\documentclass[trackchanges]{aastex701}

\usepackage{subfigure}
\usepackage{multirow}
\usepackage{longtable}
\usepackage{amsthm,amsmath,amssymb}
\usepackage{mathrsfs}
\usepackage{rotating}
\usepackage{threeparttable}
\usepackage{graphicx}

\graphicspath{{./}{fig/}}

\begin{document}

\title{Modeling Gamma-Ray Burst Spectra with Convolutional Neural Networks: Fast-Cooling Synchrotron Emission in a Decaying Magnetic Field}

\correspondingauthor{Li Zhang}\email{lizhang@ynu.edu.cn}
\author[0000-0001-5681-6939]{Jia-Ming Chen}
\affiliation{Department of Astronomy, School of Physics and Astronomy, Key Laboratory of Astroparticle Physics of Yunnan Province, Yunnan University, Kunming 650091, People’s Republic of China}
\email[show]{chenjiaming5821@163.com}

\author[0000-0002-3132-1507]{Ke-Rui Zhu}
\affiliation{Department of Astronomy, School of Physics and Astronomy, Key Laboratory of Astroparticle Physics of Yunnan Province, Yunnan University, Kunming 650091, People’s Republic of China}
\email{zhukerui0810@163.com}

\author[0000-0001-9591-4961]{Shan Chang}
\affiliation{Department of Astronomy, School of Physics and Astronomy, Key Laboratory of Astroparticle Physics of Yunnan Province, Yunnan University, Kunming 650091, People’s Republic of China}
\email{changshan@ynu.edu.cn}

\author[0000-0003-3846-0988]{Zhao-Yang Peng}
\affiliation{College of Physics and Electronics information, Yunnan Normal University, Kunming 650500, People’s Republic of China}
\email{pengzhaoyang412@163.com}

\author[0000-0003-0170-9065]{Yong-Gang Zheng}
\affiliation{Department of Physics, Yunnan Normal University, Kunming 650500, People’s Republic of China}
\email{ynzyg@ynu.edu.cn}

\author[0000-0002-5880-8497]{Li Zhang}
\affiliation{Department of Astronomy, School of Physics and Astronomy, Key Laboratory of Astroparticle Physics of Yunnan Province, Yunnan University, Kunming 650091, People’s Republic of China}
\email{lizhang@ynu.edu.cn}

\begin{abstract}
The radiation mechanism of gamma-ray burst (GRB) prompt emission remains uncertain. Although the fast-cooling synchrotron model in a decaying magnetic field can account for the characteristic nonthermal spectral shape, its computational cost has limited its use in systematic observational fitting and statistical model comparison. We develop a convolutional neural network (CNN)-based spectral emulator for this physical model and train it on a large synthetic data set generated over a physically motivated parameter space. The trained network reproduces the numerical spectra with high fidelity while reducing the cost of spectral evaluation to the millisecond level. We then incorporate the emulator into a Bayesian spectral-analysis framework and apply it to the time-resolved spectra of GRB 231020A observed by Fermi/GBM. In most time intervals, the decaying-field fast-cooling synchrotron model provides better fits and smaller Bayesian information criterion values than the standard fast-cooling synchrotron model. These results suggest that a radially decaying magnetic field provides a plausible and more physically motivated interpretation of the prompt-emission spectrum of this burst, while also indicating that the emulator offers a practical route for large-sample Bayesian inference and systematic comparisons of GRB prompt-emission models.
\end{abstract}

\section{Introduction}
\label{sec:intro}

Although studies of the prompt-emission phase of gamma-ray bursts (GRBs) have continued for decades, several key problems remain open \citep{2011CRPhy..12..206Z,2015PhR...561....1K,2018pgrb.book.....Z}. These include the jet composition, the energy-dissipation and particle-acceleration mechanisms, and the radiation mechanism itself. A reliable physical interpretation of the observed spectra is therefore central to progress on GRB prompt emission.

A typical GRB spectrum is often described by the empirical Band function \citep{1993ApJ...413..281B}. The low- and high-energy photon indices are usually around $\alpha \sim -1$ and $\beta \sim -2.2$ \citep{2000ApJS..126...19P,2006ApJS..166..298K,2021ApJ...913...60P}. Although the Band function reproduces the observed spectral shape well \citep{2009Sci...323.1688A,1999ApJ...524...82B}, its parameters are not directly tied to the underlying radiation physics. In general, prompt emission is discussed within two broad pictures. One is photospheric emission \citep{2000ApJ...530..292M,2010MNRAS.407.1033B,2017IJMPD..2630018P}, in which photons escape once the outflow becomes optically thin. Purely photospheric models, however, tend to produce thermal or quasi-thermal spectra and therefore cannot easily explain the strongly nonthermal spectra observed in most GRBs unless additional dissipation and reprocessing are involved \citep{2010ApJ...725.1137L,2011ApJ...732...49P,2013MNRAS.428.2430L}.

Another mainstream picture is that dissipation occurs within the outflow, for example through internal shocks \citep{1994ApJ...430L..93R,1998MNRAS.296..275D} or magnetic reconnection \citep{2011ApJ...726...90Z}, which accelerate nonthermal particles. These particles may then radiate through synchrotron emission \citep{2000ApJ...540..704M} or inverse Compton scattering \citep{2000ApJ...544L..17P,2008MNRAS.384...33K,2012MNRAS.424.3192B,2019ApJ...877...89Z}. Synchrotron radiation has a natural advantage in producing nonthermal spectra \citep{2000MNRAS.313L...1G,2014ApJ...784...17B,2020NatAs...4..174B,2020NatAs...4..210Z}. However, the standard fast-cooling synchrotron model with a constant magnetic field often predicts a low-energy slope softer than the typical observed value of $\alpha \sim -1$ \citep{2000ApJS..126...19P,2011A&A...530A..21N}. A physically motivated modification is to allow the magnetic field in the emitting region to decay with radius \citep{2014NatPh..10..351U,2014ApJ...780...12Z}. If the emission comes from relativistic electrons accelerated at large radii, for example in a Poynting-flux-dominated jet powered by magnetic dissipation \citep{2003astro.ph.12347L,2011ApJ...726...90Z}, the emitting region expands outward and the magnetic-field strength decreases with radius. Under this condition, the decaying-field fast-cooling synchrotron model can naturally produce Band-like spectra and has shown promising performance in GRB spectral fitting \citep{2018ApJS..234....3G}.

Despite its physical appeal, this model is computationally expensive. Previous applications have therefore been largely limited to case studies. \citet{2016ApJ...816...72Z} first applied the decaying-field synchrotron model to time-resolved spectra of GRB~130606B. \citet{2023ApJ...947L..11Y} used the same model to study the extremely bright GRB~221009A and found that its prompt spectra are consistent with synchrotron emission from relativistic electrons accelerated at large radii. \citet{2024ApJ...962...85Y} later applied a similar framework to self-consistently model the spectra of eight single-pulse GRBs. Direct numerical evaluation, however, remains impractical for systematic time-resolved fitting, broad parameter-space exploration, and large-sample statistical tests.

In recent years, machine-learning methods have been widely used in astronomy and in GRB studies \citep{2024ApJ...963...71B,2024ApJ...971...70S,2024MNRAS.527.4272C,2025ApJS..276...62C}. Here we use a convolutional neural network (CNN) to emulate the spectral calculation and remove the main computational bottleneck. Once trained, the network evaluates the spectrum for a given set of physical parameters in about $\sim 1$ ms, making repeated forward folding and likelihood evaluation feasible in routine Bayesian fitting. Building the synthetic training set is computationally expensive at the outset, but the trained emulator provides an efficient surrogate for direct comparisons between complex physical models and observational data.

This paper is organized as follows. Section~\ref{section2} presents the physical model and the synthetic data set. Section~\ref{section3} describes the CNN emulator, its training procedure, and its validation. Section~\ref{section4} applies the model to the time-resolved spectra of GRB 231020A. Section~\ref{section5} summarizes the main results and discusses the limitations of the present framework. We adopt a standard $\Lambda$CDM cosmology with $H_{0}=67.4~\mathrm{km~s^{-1}~Mpc^{-1}}$, $\Omega_{\rm M}=0.315$, and $\Omega_{\Lambda}=0.685$ \citep{2020A&A...641A...6P}.

\section{Physical Model and Synthetic Data Set}
\label{section2}
\subsection{The Synchrotron Model}
We consider a relativistic thin shell ejected from the central engine with velocity $\beta c$, and describe its evolution in the comoving time $t'$.
The shell radius is
\begin{equation}
R(t') \simeq R_0 + \beta c \Gamma t',
\end{equation}
where $R_0$ is the characteristic radius when radiation starts (or when electron injection starts), and $\Gamma$ is the bulk Lorentz factor of the shell.
For a magnetically dominated jet, or for magnetic dissipation at large radii, the emitting region expands and the comoving magnetic field decays; we adopt a power-law form for the magnetic-field evolution \citep{2001A&A...369..694S,2014NatPh..10..351U},
\begin{equation}
B'(R) = B_0' \left(\frac{R}{R_0}\right)^{-a},
\end{equation}
where $B_0'$ is the magnetic-field strength at $R_0$, and $a$ is the decay index.

The magnetic field inside the spherical shell is carried along with the matter, and electrons can be accelerated into a power-law distribution through mechanisms such as magnetic dissipation \citep{2011ApJ...726...90Z}. After injection, electrons mainly cool by synchrotron cooling and adiabatic cooling.
The electron evolution follows the continuity equation \citep{2011hea..book.....L,2014NatPh..10..351U,2014ApJ...780...12Z,2018ApJS..234....3G},
\begin{equation}
\frac{\partial N(\gamma,t')}{\partial t'}
+ \frac{\partial}{\partial \gamma}\!\left[\dot{\gamma}(\gamma,t')\,N(\gamma,t')\right]
= Q(\gamma,t'),
\end{equation}
where $Q(\gamma,t')$ is the electron injection rate in the comoving frame and we assume
\begin{equation}
Q(\gamma,t')=
\begin{cases}
Q_0\,\gamma^{-p}, & \gamma_{inj} < \gamma < \gamma_{\max},\\
0, & \text{otherwise},
\end{cases}
\end{equation}
with $Q_0$ the injection rate; here $\gamma_{inj}$ and $\gamma_{\max}$ are the minimum and maximum Lorentz factors of injected electrons.
The total cooling rate $\dot{\gamma}$ includes synchrotron and adiabatic terms,
\begin{equation}
\dot{\gamma} = \dot{\gamma}_{\rm syn} + \dot{\gamma}_{e,{\rm adi}}
= -\frac{\sigma_T {B'}^{2}}{6\pi m_e c}\,\gamma^{2}
-\frac{2}{3}\frac{\beta c \Gamma}{R}\,\gamma,
\end{equation}
where $\sigma_T$ is the Thomson cross section and $m_e$ is the electron mass.

This synchrotron model has been implemented in our previous work.
In this paper, we treat it as a numerical ``black box'': the input is a set of physical parameters and the output is the predicted photon spectrum in the observer frame.
We summarize the mapping as
\begin{equation}
N(E_{\rm obs}) = N(E_{\rm obs}; \hat{t}, B_0, a, \Gamma, \gamma_{\rm inj}, p, R_0, Q_0, z),
\end{equation}
where $\hat{t}$ is the observer-frame time since the onset of an injection episode (in seconds), $B_0$ is the magnetic-field strength at $R_0$, $a$ is the decay index, $\Gamma$ is the bulk Lorentz factor, $\gamma_{inj}$ is the minimum injected Lorentz factor, $p$ is the injection index, $R_0$ is the characteristic radius, $Q_0$ is injection rate, and $z$ is the redshift.
The goal of the CNN is to learn this mapping with high accuracy and then generate spectra consistent with the numerical model at millisecond speed.

\subsection{Construction of the Synthetic Spectral Data Set}

\begin{figure}[htbp]
\centering
\begin{minipage}[t]{\textwidth}
\centering
\includegraphics[width=\textwidth]{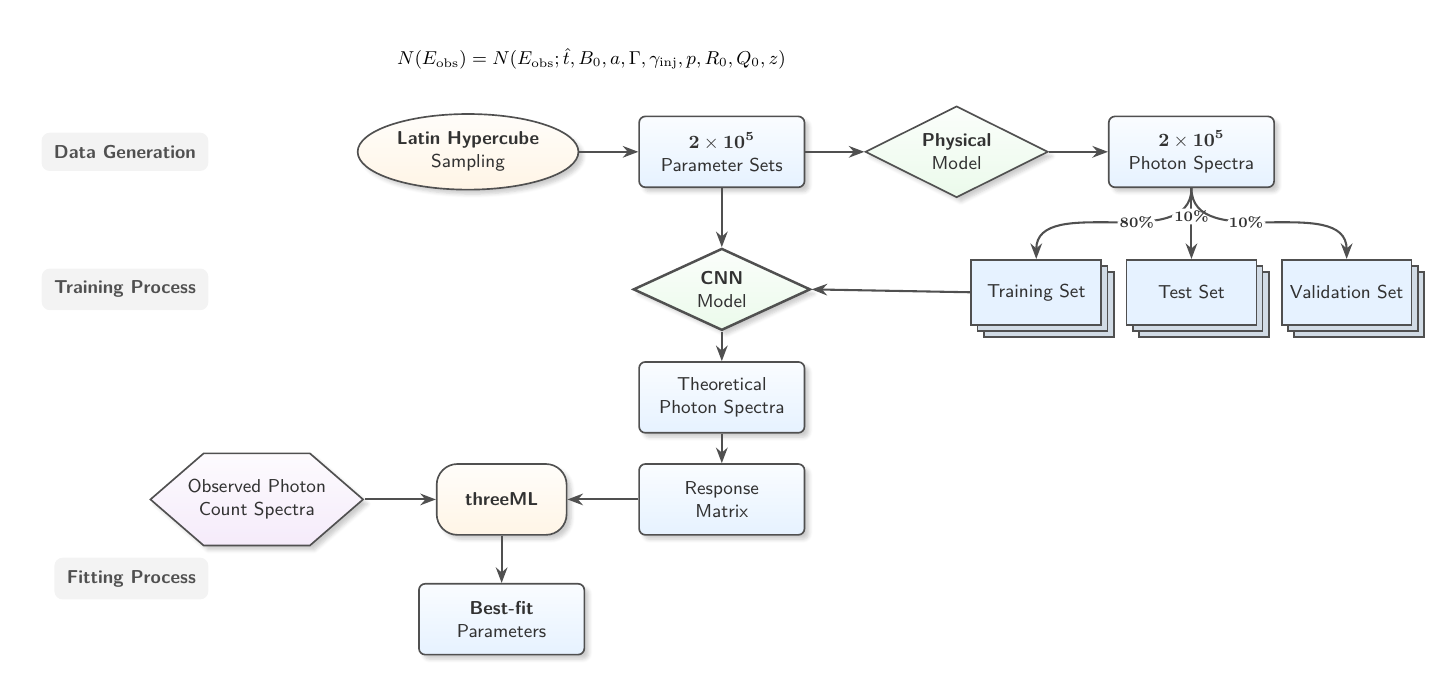}
\end{minipage}
\caption{This work proposes a workflow for GRB spectral modeling.}
\label{fig1}
\end{figure}

Figure \ref{fig1} presents the complete modeling procedure adopted in this work. In this section, we construct a synthetic spectral data set to train the CNN emulator by sampling physical parameters within reasonable ranges, calling the numerical synchrotron model to generate theoretical photon spectra, and organizing the results into supervised ``parameter--spectrum'' pairs.

The parameter vector is defined as
$\theta = (\hat{t},\, a,\, Q_0,\, \Gamma,\, R_0,\, \gamma_{inj},\, B_0,\, p,\, z)$,
with the ranges
$\hat{t} \in [0.1,\, 10]~{\rm s}$,
$a \in [0.1,\, 2.0]$,
$Q_0 \in [10^{54},\, 10^{58}]$,
$\Gamma \in [100,\, 500]$,
$R_0 \in [10^{14},\, 10^{16}]~{\rm cm}$,
$\gamma_{inj} \in [10^{4},\, 10^{6}]$,
$B_0 \in [10,\, 1000]~{\rm G}$,
$p \in [2.0,\, 5.0]$,
and $z \in [0.1,\, 10.0]$.
These intervals cover common parameter scales in GRB prompt emission while balancing physical coverage and numerical stability.

Latin hypercube sampling (LHS) is adopted to improve coverage in the high-dimensional space by first drawing $N_{\rm samp}=10^{5}$ points $\bm{u}_i$ in the unit hypercube and then mapping them to the physical parameter space; for parameters spanning multiple orders of magnitude (e.g., $Q_0$, $R_0$, and $\gamma_{inj}$), sampling is performed uniformly in log space,
$\log_{10}\theta = \log_{10}\theta_{\min} + u\left(\log_{10}\theta_{\max}-\log_{10}\theta_{\min}\right)$,
whereas for the remaining parameters a linear mapping is used,
$\theta = \theta_{\min}+u\left(\theta_{\max}-\theta_{\min}\right)$.
This strategy provides a more balanced coverage at a fixed sample size, especially for parameters that strongly affect the spectral shape.

The output spectra are defined on fixed energy grids and, to match the two \textit{Fermi}/GBM detector types, spectra are computed on two logarithmically uniform energy axes: for NaI, $E \in [8,\,10^{3}]$~keV, and for BGO, $E \in [200,\,4\times10^{4}]$~keV.
Each band uses $N_E=128$ energy bins, and for each parameter set $\bm{\theta}_j$ the photon spectrum $\bm{y}_j$ is evaluated on both the NaI and BGO grids, giving a total of $2N_{\rm samp}=2\times10^{5}$ samples. We also assign each spectrum a detector label(${\tt inst\_id}=0$ for NaI and ${\tt inst\_id}=1$ for BGO), allowing the network to distinguish between the two energy ranges.

The main computational cost arises from solving the electron evolution equation and performing the radiation integration, so multi-processing is used to generate spectra in batches.
For data-quality control, defensive processing is applied to the outputs: if a model call fails or returns a spectrum with an incorrect length, it is replaced with a tiny positive spectrum; a lower floor is imposed on all spectral values (e.g., $\ge 10^{-30}$) to avoid numerical issues in later log transforms or normalization; samples containing NaN/Inf values are removed; and the total flux is required to satisfy $\sum_i y_i > 10^{-28}$ to exclude near-zero spectra. The final data set is stored in {\tt .h5} format, containing the parameter array ({\tt float64}), the detector label ({\tt int64}), and the spectral array ({\tt float32}), together with the two energy grids and the parameter names for reproducibility and data provenance.

\section{Convolutional Neural Network Model and Training Procedure}
\label{section3}
\subsection{Data Preprocessing}

The physical parameters span very different scales, and the photon spectra vary across several orders of magnitude, so direct regression in the raw parameter space can lead to slower and less stable optimization; therefore, before training, we apply consistent and reproducible preprocessing and normalization to both inputs and outputs.
We use the parameter vector $\theta = (\hat{t},\, a,\, Q_0,\, \Gamma,\, R_0,\, \gamma_{\rm inj},\, B_{0},\, p,\, z)$, where $Q_0$, $R_0$, and $\gamma_{\rm inj}$ are strictly positive and span many decades; we first check positivity and remove invalid samples, and then transform these parameters to log space,
$Q_0 \rightarrow \log_{10} Q_0$,
$R_0 \rightarrow \log_{10} R_0$,
and $\gamma_{\rm inj} \rightarrow \log_{10} \gamma_{\rm inj}$,
while all other parameters remain in linear scale.
To reduce optimization bias caused by different units and ranges, the transformed 9D vector is standardized to zero mean and unit variance, with the mean and variance computed from the training set and reused in inference.

To enable a single network to model both the NaI and BGO bands, we encode the detector type as a one-hot vector $\mathbf{e}_{\rm inst}\in\mathbb{R}^{2}$; the final input feature is
$\mathbf{X}=[\boldsymbol{\theta}_{\rm std},\,\mathbf{e}_{\rm inst}] \in \mathbb{R}^{11}$,
so the input dimension is fixed to 11.

Output preprocessing is also important: to suppress the large dynamic range and to avoid non-numerical values caused by zeros, we first apply a lower floor to the spectrum,
$s \sim \max(s,\,10^{-30})$,
and then take a log transform,
$y_{\log}=\log_{10} s$; on this basis, $y_{\log}$ is standardized in each energy-bin dimension, yielding the training target $\mathbf{y}\in\mathbb{R}^{128}$.
This ``log + standardization'' scheme makes the network learn local slopes and curvature more effectively and improves gradient-scale consistency during training.
The output standardizer is stored for inverse transforms in inference,
$\hat{\mathbf{y}}_{\log}=\hat{\mathbf{y}} \odot \boldsymbol{\sigma}_s + \boldsymbol{\mu}_s,\ \hat{\mathbf{s}} = 10^{\hat{\mathbf{y}}_{\log}}$.

The data are randomly split into training, validation, and test sets with a fixed random seed, using a ratio of 80\%/10\%/10\%; the validation set is used for early stopping and model selection, while the test set is reserved for an independent evaluation after training.
For reproducibility, a fixed seed is also used for batch shuffling in the training set, and all preprocessing is implemented with the \texttt{Scikit-learn} package in \texttt{Python}.

\subsection{CNN Architecture}

\begin{figure}[htbp]
\centering
\begin{minipage}[t]{\textwidth}
\centering
\includegraphics[width=\textwidth]{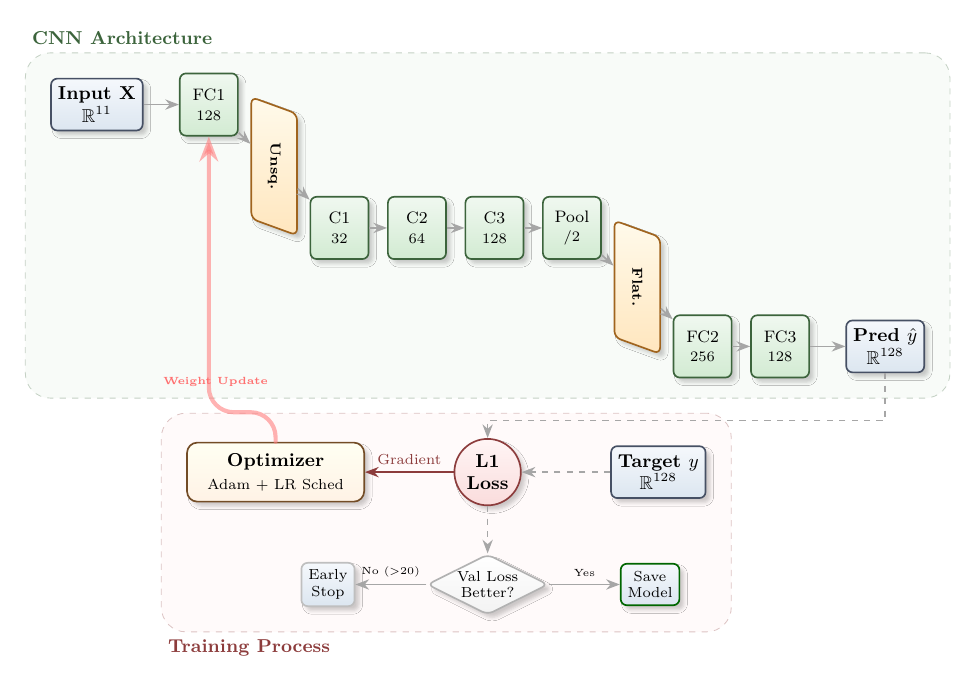}
\end{minipage}
\caption{Schematic of the CNN spectral emulator and the training workflow. }
\label{fig2}
\end{figure}

We adopt a CNN-based regression model that maps an 11-dimensional input vector to a 128-dimensional output spectrum. As shown in Figure \ref{fig2}, the model first embeds the low-dimensional parameter vector into a latent sequence representation of length 128 via a fully connected layer. It then applies one-dimensional convolutions along the ``energy-bin axis'' to extract locally correlated structures, and finally uses a fully connected regression head to output the standardized log-spectrum.

Specifically, the input $\boldsymbol{X}\in\mathbb{R}^{11}$ is projected to 128 dimensions through a linear layer followed by a ReLU activation \citep{NIPS2012_c399862d}, yielding $\boldsymbol{h}_0\in\mathbb{R}^{128}$. We reshape $\boldsymbol{h}_0$ into a one-dimensional signal of shape $(\mathrm{batch},1,128)$ and stack three 1D convolutional layers (kernel size $3$; padding $=1$ to preserve the sequence length), with channel widths $1\rightarrow 32\rightarrow 64\rightarrow 128$. Each convolution is followed by a ReLU activation to introduce nonlinearity. Given the multiscale nature of spectral smoothness and local features along the energy axis, we append a single max-pooling operation, $\mathrm{MaxPool1d}(2)$, after the convolutional block to downsample the sequence length from 128 to 64. This operation improves robustness and computational efficiency while preserving the dominant spectral morphology. The pooled feature tensor is flattened and passed through a 256-dimensional fully connected layer with ReLU activation, and the network outputs a 128-dimensional regression vector as $\hat{\boldsymbol{y}}$.

The design is motivated by the fact that, although the inputs are global physical parameters, the output photon spectrum exhibits strong local correlations along the energy axis (e.g., local slopes, curvature, and morphological changes near breaks or turnovers). One-dimensional convolutions can capture such local patterns efficiently with relatively few parameters. In addition, by augmenting the input with one-hot detector information, the network can treat the distinct energy-band coverage of NaI and BGO within a unified architecture, thereby simplifying both training and inference.

\subsection{Training and Optimization}

We implement the network in \texttt{PyTorch}\footnote{\url {https://pytorch.org/}} and train it on an NVIDIA GeForce RTX~4090 GPU; since this project targets large-sample fitting and statistical analyses, we emphasize reproducibility by fixing the random seeds for \texttt{Python}, \texttt{NumPy}, and \texttt{PyTorch}, and by enabling a deterministic computation path as much as possible.

For the loss function, the standardized log-spectrum is regressed with an $\ell_{1}$ loss \citep{Huber1992,7797130},
$\mathcal{L}=|\hat{\mathbf{y}}-\mathbf{y}|_{1}$, which is more robust to local outliers than MSE, reduces the influence of a small number of extreme spectral shapes, and stabilizes the fit to the overall spectral form.

We use the Adam optimizer \citep{kingma2017adammethodstochasticoptimization} with an initial learning rate of $10^{-3}$ and adopt a step-wise learning-rate decay (\texttt{StepLR}) to realize a ``fast convergence then fine tuning'' schedule; every 50 epochs the learning rate is multiplied by 0.1.
The maximum number of epochs is set to 2500, while the effective training length is determined by early stopping based on the validation loss: if no significant improvement is found for 20 consecutive epochs (improvement $<10^{-4}$), training is terminated and the model weights are restored to those at the epoch with the best validation loss, so that the final network is selected by generalization performance rather than by a fixed epoch number.

\subsection{Training Results}

\begin{figure}[htbp]
\centering
\begin{minipage}[t]{\textwidth}
\centering
\includegraphics[width=0.6\textwidth]{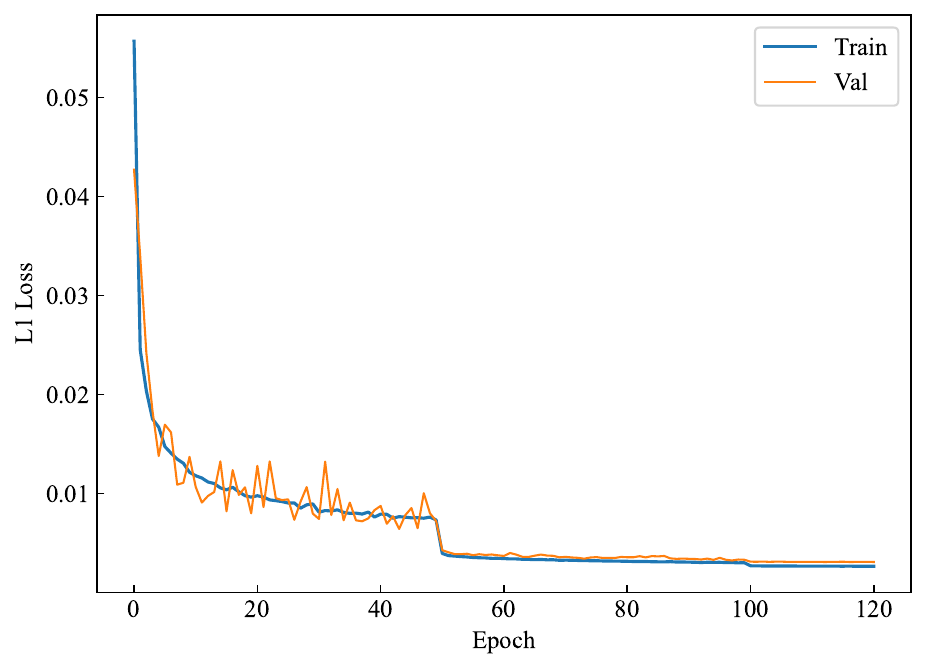}
\end{minipage}
\caption{Training and validation $\ell_1$ loss as a function of epoch for the CNN spectral emulator. Both curves decrease rapidly during the initial epochs and then gradually converge, indicating stable optimization and good generalization with no evidence for overfitting.}
\label{fig3}
\end{figure}

We train the CNN model following the strategy described above, and Figure \ref{fig3} shows the evolution of the $L_1$ loss on the training and validation sets as a function of epoch. The loss drops rapidly in the first few epochs, indicating that the network quickly learns the dominant spectral features controlled by the physical parameters and reduces the overall prediction error efficiently. As training proceeds, the decrease of the loss slows down and the curves enter a plateau; the training and validation losses remain close and follow the same trend, with only small random fluctuations at early epochs, suggesting a stable optimization process without obvious overfitting. We also note a clear step-like decrease of the loss around epoch $\sim 50$, consistent with the step-wise learning-rate decay (\texttt{StepLR}); after the learning rate is reduced, the model further refines the fit based on the learned representation and converges at a lower loss level.

\begin{figure}[htbp]
\centering
\begin{minipage}[t]{\textwidth}
\centering
\includegraphics[width=0.8\textwidth]{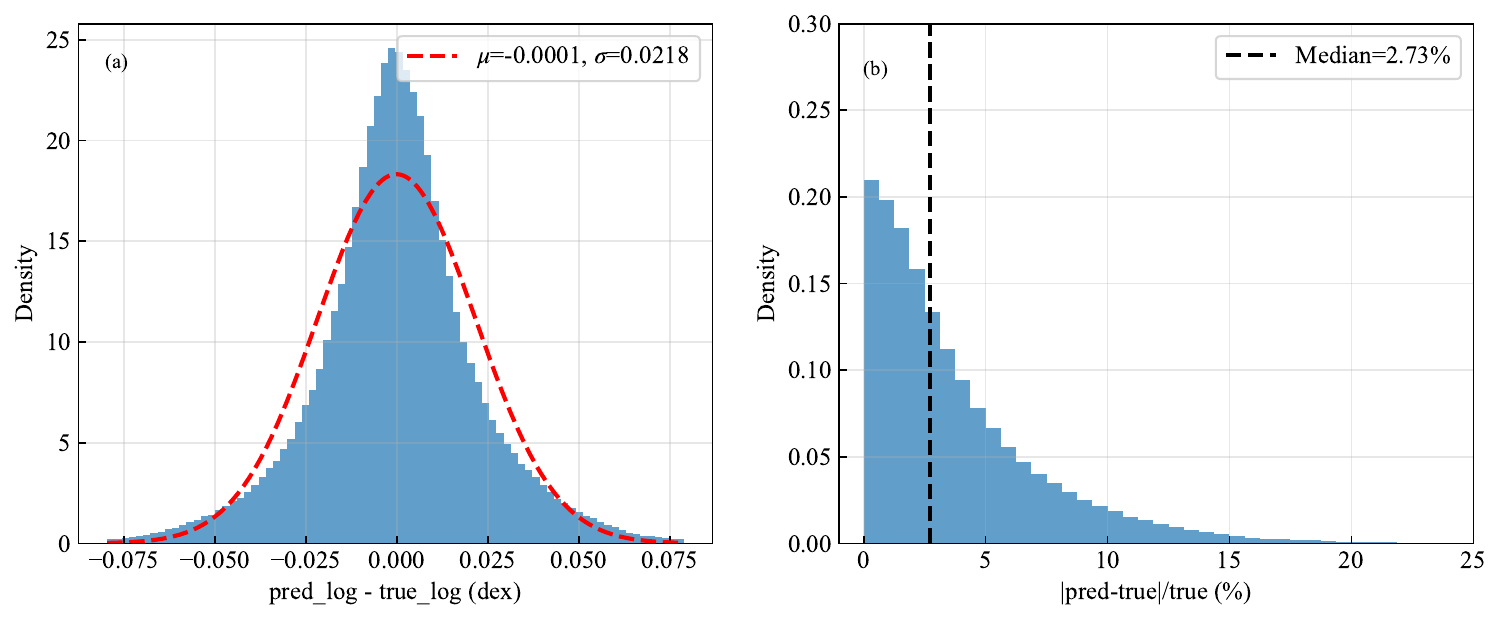}
\end{minipage}
\caption{Validation statistics of the CNN spectral emulator on the independent test set. Left: distribution of logarithmic residuals, defined as $\log_{10}(F_{\rm CNN}/F_{\rm num})$, showing negligible systematic bias and a narrow scatter around zero. Right: distribution of relative errors, with a median of $\sim 2.73\%$ and a 95th-percentile value of $\sim 11.22\%$.
}
\label{fig4}
\end{figure}

We further evaluate the prediction errors on an independent test set. Figure \ref{fig4}(a) shows the histogram of log-residuals, with a mean close to zero ($\mu \simeq -1.0\times10^{-4}$~dex), implying little systematic bias, and a standard deviation of $\sigma \simeq 2.18\times10^{-2}$~dex, corresponding to a typical scatter of $\sim 5.0\%$ in linear space and indicating that the target spectra are reproduced with high accuracy. Figure \ref{fig4}(b) summarizes the relative error statistics: the mean relative error is $\simeq 3.88\%$, the median is $\simeq 2.73\%$, and the 95th-percentile error is $\simeq 11.22\%$; the distribution is right-skewed with a long tail, meaning that most spectra are reconstructed within a few percent while a small fraction of difficult samples contributes to the tail, usually associated with weak-flux energy bins or strong spectral curvature.

\begin{figure}[htbp]
\centering
\begin{minipage}[t]{\textwidth}
\centering
\includegraphics[width=\textwidth]{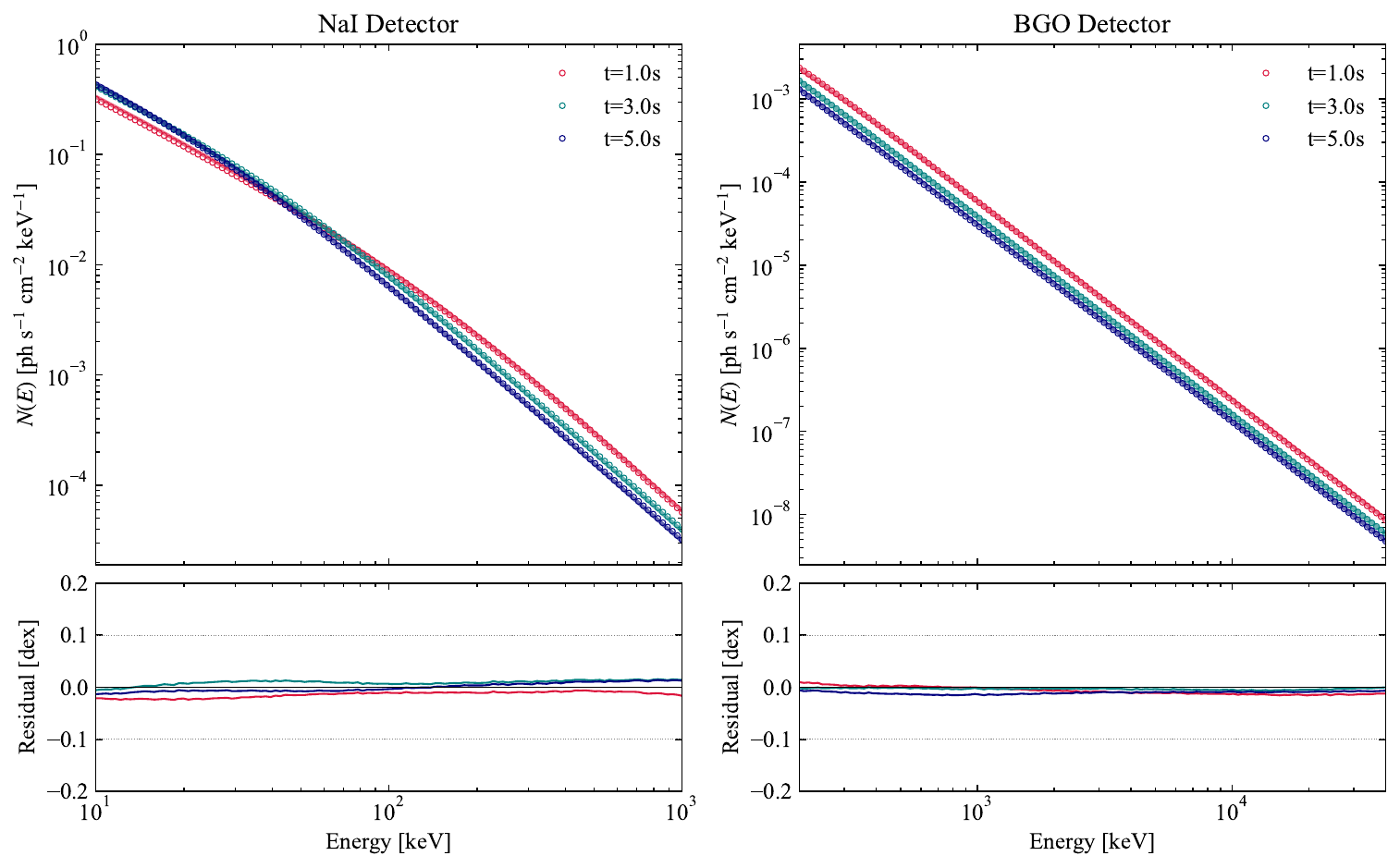}
\end{minipage}
\caption{Comparison between the CNN-emulated photon spectra and the corresponding numerical-model spectra on the fixed \textit{Fermi}/GBM energy grids. The left and right columns show the NaI and BGO bands, respectively, for three representative epochs ($t=1.0,\,3.0,$ and $5.0$~s; colors). The upper panels present $N(E)$ as a function of energy, while the lower panels show the logarithmic residuals (in dex) between the emulator prediction and the numerical reference. Solid curves and open circles denote the numerical-model spectra and the CNN-emulated photon spectra, respectively.}
\label{fig5}
\end{figure}

Overall, the training converges fast and remains stable, with consistent performance between the training and validation sets, while the test set shows near-zero bias and small scatter, supporting good generalization. Figure \ref{fig5} compares spectra from the numerical physical model and from the CNN emulator, showing good agreement across the full energy band; the CNN reproduces both spectral slopes and curvature features, and provides millisecond-level inference that yields spectra consistent with the numerical synchrotron model. This greatly reduces the computing time for large-scale parameter scans and spectral fitting, and provides a reliable basis for subsequent large-sample fitting, parameter-space exploration, and statistical analyses.

\section{Time-resolved Spectral Constraints on GRB 231020A}
\label{section4}

\subsection{Data selection and spectral fitting}
\label{subsec:data_fitting}

To demonstrate the efficiency and practical utility of our CNN-based spectral emulator in real data fitting, we perform a time-resolved spectral analysis of a GRB observed by \textit{Fermi}. For seamless use in an observational fitting pipeline, we package the trained CNN emulator as a custom spectral model through the user-defined model interface provided by \texttt{threeML} (3ML, \citet{2015arXiv150708343V}). The \texttt{threeML} framework is widely used for \textit{Fermi}/GBM GRB spectral analyses, and all spectral analyses in this work are carried out within 3ML \citep{2019ApJ...886...20Y,2022ApJ...932...25C,2024ApJ...972..132C}.

GRB~231020A (Trigger 719521023/231020790) triggered the \textit{Fermi} Gamma-ray Burst Monitor (GBM) and was localized at 2023 October 20 18:56:58.30 UT \citep{2023GCN.34856....1F}. According to the \textit{Fermi} team report, the light curve is dominated by a bright main pulse with a duration of $T_{90}\approx 9.3$~s in the 50--300~keV band. For the time-integrated spectrum from $T_0-0.5$~s to $T_0+21$~s, the best-fit model is a cutoff power law (CPL), with a low-energy index $\alpha=-1.34\pm0.01$ and a cutoff energy $E_{\rm c}=145\pm3$~keV. The relatively soft low-energy slope and the pronounced non-thermal spectral shape suggest a synchrotron origin, making this burst a suitable test case for our fast-cooling synchrotron model in a decaying magnetic field and its CNN emulator.

Based on detector viewing angles and count-rate significance, we select three NaI detectors (n2, n9, na) and one BGO detector (b1) for the spectral analysis. For background estimation, we use the photon counts of the brightest NaI detector as the reference and choose two off-source intervals, one before and one after the burst. We fit the background with polynomials of order 0--4, determine the optimal order using a likelihood-ratio test, and extrapolate the selected background model into the source interval. We then subtract the background from the counts in all 128 energy channels.

\begin{figure}[htbp]
\centering
\includegraphics[width=0.6\textwidth]{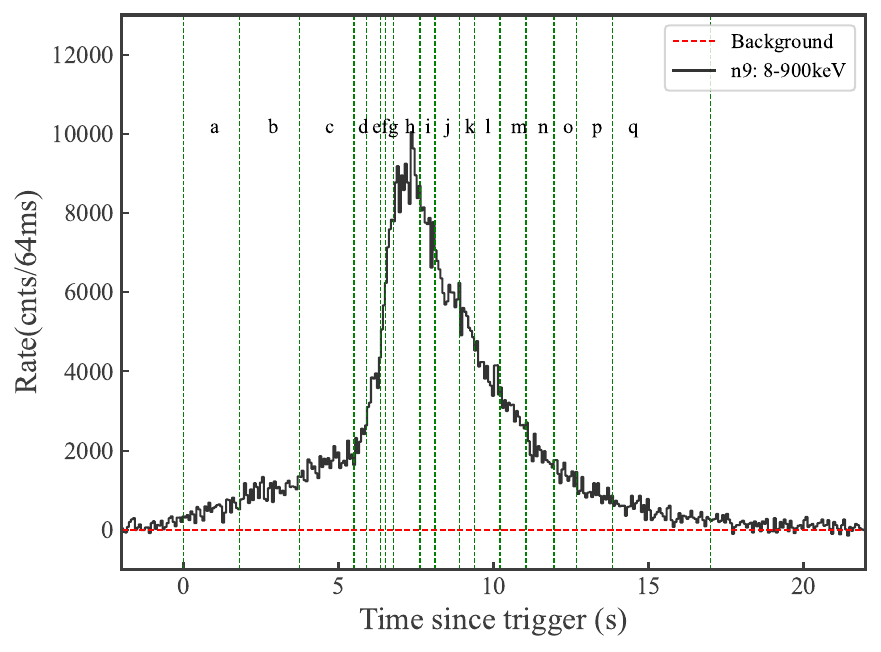}
\caption{Light curve of GRB~231020A. The green dashed vertical lines mark the Bayesian Blocks change points, defining the time intervals used for the time-resolved spectral analysis.}
\label{fig6}
\end{figure}

Time-resolved spectroscopy requires a sufficiently fine partition of the light curve to track spectral evolution. We therefore apply the Bayesian blocks method to the TTE light curve of the brightest NaI detector, adopting a false-alarm probability of $p=0.01$ \citep{2013ApJ...764..167S}. As shown in Figure~\ref{fig6}, this adaptive segmentation yields a total of 18 time bins, which are used for the subsequent time-resolved spectral fitting.

After defining the time-resolved bins, we perform Bayesian spectral fitting for each time interval in 3ML using the packaged CNN synchrotron model, denoted as \texttt{ME\_FCSYN\_CNN}. To ensure that the inference remains within the training coverage of the CNN, we set the parameter priors to be consistent with the ranges used in the synthetic data set. For parameters that span multiple orders of magnitude, we adopt a logarithmic parametrization to improve sampling efficiency. Specifically, we use $\hat{t}\sim \mathcal{U}(0.1,\,10.0)$, $a\sim \mathcal{N}(1.0,\,0.5)$ truncated to $[0.1,\,2.0]$, $\log_{10}\gamma_{\rm inj}\sim \mathcal{U}(4,\,6)$, $\log_{10}\Gamma\sim \mathcal{U}(2,\,2.7)$, $\log_{10}Q_0\sim \mathcal{U}(54,\,58)$, $\log_{10}B_{0}\sim \mathcal{U}(1,\,3)$, $\log_{10}R_0\sim \mathcal{U}(14,\,16)$, and $p\sim \mathcal{N}(2.8,\,1)$ truncated to $[2.2,\,5.0]$. Here $\hat{t}$, $a$, and $p$ are sampled in linear space, while $Q_0$, $R_0$, $\gamma_{\rm inj}$, $\Gamma$, and $B_{0}$ are sampled in log space. Since no redshift measurement is available for GRB~231020A, we fix the redshift to $z=1$.

We then construct a \texttt{BayesianAnalysis} object in 3ML and use \texttt{MultiNest} as the sampler. Given the relatively high parameter dimension and potential parameter correlations, the number of live points is set to $n_{\rm live}=1000$ to improve the stability of posterior sampling and the accuracy of the evidence estimate.

For comparison with a physical baseline, we also fit each time bin with a standard fast-cooling synchrotron model, denoted as \texttt{FCSYN}. The model form and implementation follow the treatments in \citet{2014ApJ...784...17B} and \citet{2025ApJ...993..147C}. For the standard \texttt{FCSYN} model, we fix $\gamma_{\rm inj}=10^5$ and $\gamma_{\rm cool}=10^2$ to maintain the fast-cooling regime \citep{2020NatAs...4..174B}. We note that when both $B$ and $\gamma_{\rm inj}$ are allowed to vary freely, the two parameters become strongly degenerate because the synchrotron peak energy is mainly controlled by their combination. Therefore, the FCSYN magnetic-field parameter should be regarded as an effective constant-field value and should not be directly identified with $B_0$ in the decaying-field model. The \texttt{FCSYN} model is used here only as a simple standard fast-cooling synchrotron comparison model.  We adopt the Bayesian Information Criterion (BIC) and the Residual Sum of Squares (RSS) as quantitative metrics for model selection and goodness of fit.

\begin{figure}[htbp]
\centering
\includegraphics[width=\textwidth]{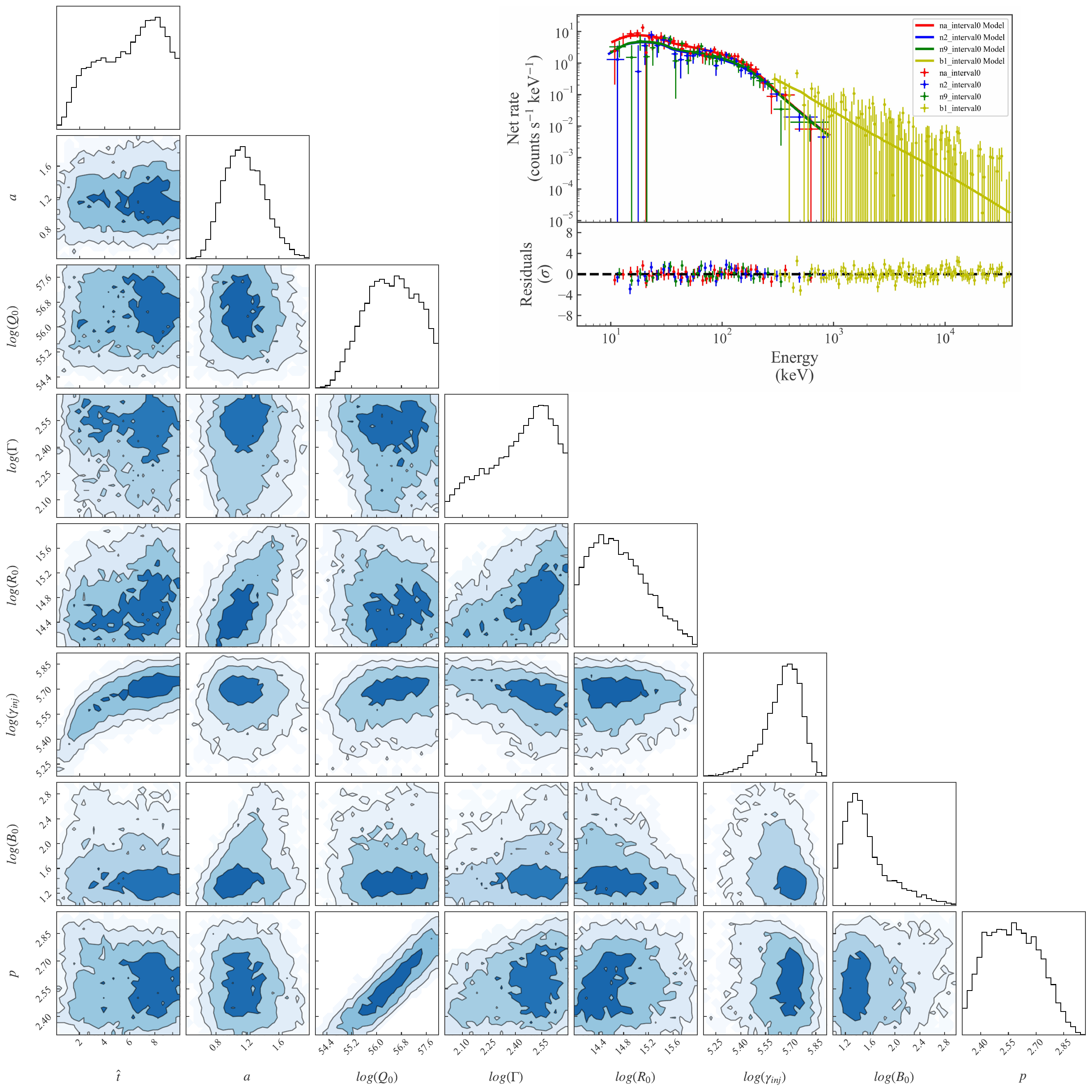}
\caption{GRB~231020A: representative fitting result for one time bin. Left: corner plot of the posterior distributions for the \texttt{ME\_FCSYN\_CNN} model parameters from the Bayesian fit. Right: best-fit forward-folded GBM count spectra and residuals for the same interval.}
\label{fig:posterior_and_fit}
\end{figure}

We apply both \texttt{ME\_FCSYN\_CNN} and \texttt{FCSYN} to the time-resolved spectra of GRB~231020A. The fitting results for all bins are summarized in Table~\ref{tab:results}. As an example, Figure~\ref{fig:posterior_and_fit} shows the fitted count spectrum and the posterior distributions of the model parameters for one representative time bin. From the residual behavior and the RSS values, \texttt{ME\_FCSYN\_CNN} provides acceptable fits to the data across the GBM energy range.

\begin{longrotatetable}
\setlength{\tabcolsep}{2.5pt} %
\begin{deluxetable}{llcccccccccc}
\tablewidth{0pt}
\tablecaption{Time-resolved spectral-fitting results for GRB 231020A.\label{tab:results}}

\tablehead{
\colhead{Time[s]} & \colhead{Model} & \colhead{$\hat{t}[s]$} & \colhead{$b/\alpha$} & \colhead{$\log Q_0[s^{-1}]$} & \colhead{$\log\Gamma$} & \colhead{$\log R_0[cm]$} & \colhead{$\log B_0/B[G]$} & \colhead{$p/\beta$} & \colhead{$E_{\rm p}[keV]$} & \colhead{$\mathrm{RSS}/\mathrm{dof}$} & \colhead{BIC}
}

\startdata
0.00$\sim$1.82 & FCSYN & -- & -- & -- & -- & -- & $17697.36^{+2867.68}_{-3139.23}$ & $2.31^{+0.86}_{-0.24}$ & -- & 431.03/473 & 3087.18 \\
 & ME\_FCSYN\_CNN & $9.89^{+1.04}_{-6.23}$ & $1.31^{+0.12}_{-0.38}$ & $57.32^{+0.17}_{-1.49}$ & $2.65^{+0.07}_{-0.49}$ & $14.33^{+0.72}_{-0.12}$ & $2.46^{+0.43}_{-1.26}$ & $2.65^{+0.05}_{-0.20}$ & -- & 415.58/468 & 3064.82 \\
 & Band & -- & $-0.69^{+0.11}_{-0.14}$ & -- & -- & -- & -- & $-2.10^{+0.01}_{-0.61}$ & $251.77^{+72.57}_{-22.87}$ & 412.65/472 & 3066.70 \\
 & CPL & -- & $-0.76^{+0.12}_{-0.11}$ & -- & -- & -- & -- & -- & $306.48^{+61.31}_{-40.34}$ & 417.26/473 & 3065.50 \\
\hline
1.82$\sim$3.74 & FCSYN & -- & -- & -- & -- & -- & $10.00^{+0.91}_{-0.08}$ & $2.72^{+0.15}_{-0.01}$ & -- & 422.16/473 & 3191.93 \\
 & ME\_FCSYN\_CNN & $7.08^{+0.56}_{-1.01}$ & $1.11^{+0.11}_{-0.59}$ & $57.32^{+0.68}_{-2.72}$ & $2.39^{+0.09}_{-0.28}$ & $14.56^{+1.25}_{-0.20}$ & $1.72^{+0.83}_{-0.35}$ & $2.76^{+0.02}_{-0.34}$ & -- & 405.79/468 & 3163.36 \\
 & Band & -- & $-1.12^{+0.24}_{-0.23}$ & -- & -- & -- & -- & $-2.00^{+0.01}_{-0.09}$ & $52.75^{+18.98}_{-7.23}$ & 404.55/472 & 3170.77 \\
 & CPL & -- & $-1.61^{+0.04}_{-0.08}$ & -- & -- & -- & -- & -- & $155.79^{+67.59}_{-18.79}$ & 416.37/473 & 3177.45 \\
\hline
3.74$\sim$5.50 & FCSYN & -- & -- & -- & -- & -- & $10.00^{+0.13}_{-0.01}$ & $2.84^{+0.00}_{-0.03}$ & -- & 785.15/473 & 3439.03 \\
 & ME\_FCSYN\_CNN & $0.45^{+0.65}_{-0.27}$ & $1.19^{+0.10}_{-0.51}$ & $55.62^{+0.03}_{-1.26}$ & $2.05^{+0.34}_{-0.02}$ & $15.13^{+0.63}_{-0.36}$ & $1.08^{+1.15}_{-0.18}$ & $2.44^{+0.12}_{-0.10}$ & -- & 435.68/468 & 3063.58 \\
 & Band & -- & $-1.14^{+0.29}_{-0.12}$ & -- & -- & -- & -- & $-2.13^{+0.03}_{-0.03}$ & $19.92^{+1.54}_{-1.55}$ & 431.09/472 & 3085.14 \\
 & CPL & -- & $-1.97^{+0.03}_{-0.01}$ & -- & -- & -- & -- & -- & $22.63^{+25.93}_{-1.14}$ & 450.29/473 & 3101.94 \\
\hline
5.50$\sim$5.91 & FCSYN & -- & -- & -- & -- & -- & $10.01^{+0.32}_{-0.03}$ & $2.82^{+0.00}_{-0.05}$ & -- & 407.40/473 & 1381.24 \\
 & ME\_FCSYN\_CNN & $0.85^{+0.32}_{-0.16}$ & $1.05^{+0.20}_{-0.45}$ & $55.72^{+0.82}_{-0.83}$ & $2.17^{+0.33}_{-0.06}$ & $15.74^{+0.03}_{-1.07}$ & $1.14^{+1.47}_{-0.39}$ & $2.52^{+0.26}_{-0.07}$ & -- & 303.06/468 & 1208.03 \\
 & Band & -- & $-1.42^{+0.38}_{-0.00}$ & -- & -- & -- & -- & $-2.21^{+0.05}_{-0.07}$ & $24.75^{+3.65}_{-3.84}$ & 302.55/472 & 1228.63 \\
 & CPL & -- & $-1.88^{+0.09}_{-0.04}$ & -- & -- & -- & -- & -- & $35.30^{+13.58}_{-4.71}$ & 314.03/473 & 1225.22 \\
\hline
5.91$\sim$6.35 & FCSYN & -- & -- & -- & -- & -- & $10.00^{+0.19}_{-0.01}$ & $2.77^{+0.00}_{-0.03}$ & -- & 440.93/473 & 1569.22 \\
 & ME\_FCSYN\_CNN & $8.98^{+0.66}_{-6.65}$ & $1.15^{+0.10}_{-0.63}$ & $57.62^{+0.23}_{-2.20}$ & $2.10^{+0.39}_{-0.01}$ & $15.64^{+0.16}_{-0.81}$ & $2.96^{+0.20}_{-0.92}$ & $2.94^{+0.01}_{-0.41}$ & -- & 325.60/468 & 1387.74 \\
 & Band & -- & $-1.50^{+0.26}_{-0.02}$ & -- & -- & -- & -- & $-2.22^{+0.11}_{-0.08}$ & $47.67^{+3.80}_{-14.40}$ & 324.01/472 & 1409.21 \\
 & CPL & -- & $-1.76^{+0.06}_{-0.05}$ & -- & -- & -- & -- & -- & $73.23^{+10.54}_{-6.83}$ & 330.28/473 & 1396.02 \\
\hline
6.35$\sim$6.52 & FCSYN & -- & -- & -- & -- & -- & $10.00^{+0.72}_{-0.07}$ & $2.73^{+0.01}_{-0.09}$ & -- & 307.64/473 & 650.91 \\
 & ME\_FCSYN\_CNN & $5.68^{+2.51}_{-2.23}$ & $0.98^{+0.24}_{-0.45}$ & $57.75^{+0.04}_{-1.55}$ & $2.44^{+0.14}_{-0.28}$ & $15.46^{+0.34}_{-0.55}$ & $1.75^{+0.96}_{-0.04}$ & $2.91^{+0.05}_{-0.30}$ & -- & 313.80/468 & 608.31 \\
 & Band & -- & $-1.50^{+0.10}_{-0.01}$ & -- & -- & -- & -- & $-2.71^{+0.32}_{-0.37}$ & $121.37^{+8.45}_{-19.42}$ & 318.92/472 & 623.88 \\
 & CPL & -- & $-1.53^{+0.07}_{-0.06}$ & -- & -- & -- & -- & -- & $131.28^{+20.92}_{-14.11}$ & 320.77/473 & 616.22 \\
\hline
6.52$\sim$6.78 & FCSYN & -- & -- & -- & -- & -- & $10.01^{+1.50}_{-0.19}$ & $2.63^{+0.08}_{-0.02}$ & -- & 337.85/473 & 1118.97 \\
 & ME\_FCSYN\_CNN & $1.31^{+6.79}_{-0.64}$ & $1.06^{+0.19}_{-0.44}$ & $57.69^{+0.15}_{-0.69}$ & $2.67^{+0.04}_{-0.41}$ & $15.62^{+0.13}_{-0.97}$ & $1.96^{+0.52}_{-0.29}$ & $2.99^{+0.07}_{-0.27}$ & -- & 336.16/468 & 1084.99 \\
 & Band & -- & $-1.43^{+0.03}_{-0.04}$ & -- & -- & -- & -- & $-2.96^{+0.28}_{-0.33}$ & $177.21^{+17.43}_{-13.50}$ & 338.54/472 & 1102.70 \\
 & CPL & -- & $-1.44^{+0.04}_{-0.04}$ & -- & -- & -- & -- & -- & $181.76^{+17.03}_{-13.25}$ & 339.67/473 & 1089.74 \\
\hline
6.78$\sim$7.63 & FCSYN & -- & -- & -- & -- & -- & $24.32^{+1.52}_{-1.52}$ & $2.75^{+0.00}_{-0.05}$ & -- & 875.58/473 & 3016.53 \\
 & ME\_FCSYN\_CNN & $0.11^{+0.08}_{-0.00}$ & $0.92^{+0.23}_{-0.09}$ & $57.99^{+0.01}_{-0.20}$ & $2.69^{+0.00}_{-0.05}$ & $14.01^{+1.18}_{-0.04}$ & $2.47^{+0.03}_{-0.90}$ & $3.02^{+0.02}_{-0.07}$ & -- & 577.23/468 & 2652.64 \\
 & Band & -- & $-1.19^{+0.02}_{-0.02}$ & -- & -- & -- & -- & $-3.50^{+0.20}_{-0.33}$ & $219.60^{+8.12}_{-7.41}$ & 516.47/472 & 2614.96 \\
 & CPL & -- & $-1.19^{+0.02}_{-0.02}$ & -- & -- & -- & -- & -- & $221.65^{+8.09}_{-6.94}$ & 513.46/473 & 2582.55 \\
\hline
7.63$\sim$8.11 & FCSYN & -- & -- & -- & -- & -- & $27.55^{+2.32}_{-2.83}$ & $2.76^{+0.01}_{-0.10}$ & -- & 742.83/473 & 2213.59 \\
 & ME\_FCSYN\_CNN & $0.10^{+0.10}_{-0.01}$ & $0.77^{+0.05}_{-0.06}$ & $57.99^{+0.01}_{-0.16}$ & $2.70^{+0.01}_{-0.05}$ & $14.55^{+0.02}_{-0.29}$ & $1.65^{+0.11}_{-0.01}$ & $2.99^{+0.02}_{-0.06}$ & -- & 446.80/468 & 1859.24 \\
 & Band & -- & $-0.95^{+0.04}_{-0.03}$ & -- & -- & -- & -- & $-3.04^{+0.15}_{-0.32}$ & $173.16^{+8.68}_{-6.28}$ & 410.69/472 & 1819.24 \\
 & CPL & -- & $-0.97^{+0.03}_{-0.03}$ & -- & -- & -- & -- & -- & $181.11^{+7.48}_{-6.72}$ & 419.46/473 & 1806.51 \\
\hline
8.11$\sim$8.90 & FCSYN & -- & -- & -- & -- & -- & $29.34^{+1.92}_{-2.77}$ & $2.79^{+0.01}_{-0.07}$ & -- & 1205.14/473 & 3177.50 \\
 & ME\_FCSYN\_CNN & $0.14^{+9.77}_{-0.08}$ & $0.69^{+0.03}_{-0.09}$ & $57.91^{+0.08}_{-0.02}$ & $2.66^{+0.04}_{-0.00}$ & $14.17^{+0.04}_{-0.10}$ & $1.67^{+0.04}_{-0.19}$ & $2.94^{+0.02}_{-0.11}$ & -- & 615.46/468 & 2536.44 \\
 & Band & -- & $-0.69^{+0.03}_{-0.05}$ & -- & -- & -- & -- & $-3.03^{+0.14}_{-0.24}$ & $147.00^{+5.88}_{-4.34}$ & 426.86/472 & 2341.93 \\
 & CPL & -- & $-0.75^{+0.03}_{-0.04}$ & -- & -- & -- & -- & -- & $156.86^{+4.91}_{-3.87}$ & 444.34/473 & 2333.41 \\
\hline
8.90$\sim$9.39 & FCSYN & -- & -- & -- & -- & -- & $13.84^{+1.37}_{-1.21}$ & $2.76^{+0.01}_{-0.09}$ & -- & 692.95/473 & 1999.52 \\
 & ME\_FCSYN\_CNN & $9.98^{+0.19}_{-2.31}$ & $0.59^{+0.16}_{-0.04}$ & $58.00^{+0.02}_{-0.15}$ & $2.68^{+0.01}_{-0.05}$ & $14.21^{+0.12}_{-0.16}$ & $1.58^{+0.35}_{-0.07}$ & $2.92^{+0.01}_{-0.05}$ & -- & 479.30/468 & 1751.31 \\
 & Band & -- & $-0.62^{+0.06}_{-0.06}$ & -- & -- & -- & -- & $-3.22^{+0.15}_{-0.31}$ & $109.19^{+4.15}_{-3.51}$ & 347.77/472 & 1615.93 \\
 & CPL & -- & $-0.67^{+0.06}_{-0.05}$ & -- & -- & -- & -- & -- & $114.34^{+3.91}_{-3.57}$ & 358.28/473 & 1595.26 \\
\hline
9.39$\sim$10.21 & FCSYN & -- & -- & -- & -- & -- & $10.01^{+0.45}_{-0.04}$ & $2.77^{+0.00}_{-0.04}$ & -- & 710.72/473 & 2529.78 \\
 & ME\_FCSYN\_CNN & $9.89^{+0.42}_{-0.65}$ & $0.60^{+0.24}_{-0.05}$ & $57.99^{+0.01}_{-0.14}$ & $2.67^{+0.02}_{-0.06}$ & $14.18^{+0.82}_{-0.08}$ & $1.42^{+0.19}_{-0.06}$ & $2.91^{+0.07}_{-0.02}$ & -- & 538.72/468 & 2328.93 \\
 & Band & -- & $-0.67^{+0.06}_{-0.07}$ & -- & -- & -- & -- & $-2.98^{+0.10}_{-0.23}$ & $86.30^{+3.56}_{-2.45}$ & 392.80/472 & 2192.64 \\
 & CPL & -- & $-0.76^{+0.06}_{-0.05}$ & -- & -- & -- & -- & -- & $93.11^{+2.71}_{-2.64}$ & 411.44/473 & 2187.17 \\
\hline
10.21$\sim$11.05 & FCSYN & -- & -- & -- & -- & -- & $10.00^{+0.16}_{-0.01}$ & $2.79^{+0.00}_{-0.02}$ & -- & 602.18/473 & 2383.21 \\
 & ME\_FCSYN\_CNN & $6.49^{+1.94}_{-4.38}$ & $0.67^{+0.66}_{-0.05}$ & $57.96^{+0.01}_{-0.23}$ & $2.70^{+0.01}_{-0.16}$ & $15.23^{+0.61}_{-0.36}$ & $1.34^{+0.25}_{-0.00}$ & $3.06^{+0.00}_{-0.08}$ & -- & 439.95/468 & 2164.87 \\
 & Band & -- & $-0.91^{+0.10}_{-0.06}$ & -- & -- & -- & -- & $-3.06^{+0.16}_{-0.26}$ & $65.84^{+2.35}_{-3.22}$ & 394.46/472 & 2125.72 \\
 & CPL & -- & $-0.99^{+0.08}_{-0.06}$ & -- & -- & -- & -- & -- & $69.52^{+2.53}_{-2.06}$ & 404.68/473 & 2109.59 \\
\hline
11.05$\sim$11.95 & FCSYN & -- & -- & -- & -- & -- & $10.00^{+0.19}_{-0.01}$ & $2.83^{+0.00}_{-0.03}$ & -- & 550.29/473 & 2363.69 \\
 & ME\_FCSYN\_CNN & $1.83^{+6.51}_{-0.56}$ & $0.81^{+0.31}_{-0.22}$ & $57.97^{+0.04}_{-0.51}$ & $2.69^{+0.02}_{-0.26}$ & $15.75^{+0.06}_{-0.82}$ & $1.46^{+0.64}_{-0.02}$ & $3.21^{+0.06}_{-0.20}$ & -- & 382.57/468 & 2133.65 \\
 & Band & -- & $-1.17^{+0.13}_{-0.08}$ & -- & -- & -- & -- & $-3.18^{+0.25}_{-0.31}$ & $53.88^{+2.21}_{-3.58}$ & 371.78/472 & 2133.07 \\
 & CPL & -- & $-1.19^{+0.10}_{-0.08}$ & -- & -- & -- & -- & -- & $54.84^{+2.75}_{-2.20}$ & 374.74/473 & 2113.19 \\
\hline
11.95$\sim$12.68 & FCSYN & -- & -- & -- & -- & -- & $10.01^{+0.31}_{-0.02}$ & $2.86^{+0.00}_{-0.05}$ & -- & 496.81/473 & 1971.97 \\
 & ME\_FCSYN\_CNN & $4.57^{+3.43}_{-2.53}$ & $0.93^{+0.15}_{-0.48}$ & $57.88^{+0.03}_{-0.86}$ & $2.69^{+0.06}_{-0.47}$ & $15.99^{+0.17}_{-1.07}$ & $1.71^{+0.88}_{-0.04}$ & $3.26^{+0.04}_{-0.28}$ & -- & 377.22/468 & 1803.52 \\
 & Band & -- & $-1.22^{+0.19}_{-0.11}$ & -- & -- & -- & -- & $-3.02^{+0.24}_{-0.29}$ & $44.52^{+2.63}_{-3.97}$ & 371.88/472 & 1810.83 \\
 & CPL & -- & $-1.25^{+0.13}_{-0.12}$ & -- & -- & -- & -- & -- & $46.07^{+3.20}_{-2.81}$ & 374.87/473 & 1796.45 \\
\hline
12.68$\sim$13.83 & FCSYN & -- & -- & -- & -- & -- & $10.01^{+0.28}_{-0.02}$ & $2.89^{+0.00}_{-0.05}$ & -- & 537.13/473 & 2560.26 \\
 & ME\_FCSYN\_CNN & $1.68^{+5.98}_{-0.02}$ & $0.29^{+0.75}_{-0.05}$ & $57.91^{+0.00}_{-0.63}$ & $2.23^{+0.40}_{-0.03}$ & $15.93^{+0.05}_{-0.92}$ & $1.01^{+1.56}_{-0.50}$ & $3.11^{+0.19}_{-0.06}$ & -- & 367.21/468 & 2355.79 \\
 & Band & -- & $-0.97^{+0.28}_{-0.09}$ & -- & -- & -- & -- & $-3.06^{+0.18}_{-0.31}$ & $36.63^{+2.19}_{-2.31}$ & 352.25/472 & 2351.83 \\
 & CPL & -- & $-1.02^{+0.19}_{-0.12}$ & -- & -- & -- & -- & -- & $38.65^{+2.02}_{-2.24}$ & 354.96/473 & 2336.87 \\
\hline
13.83$\sim$15.00 & FCSYN & -- & -- & -- & -- & -- & $10.00^{+0.49}_{-0.05}$ & $2.93^{+0.00}_{-0.09}$ & -- & 509.01/473 & 2488.23 \\
 & ME\_FCSYN\_CNN & $3.11^{+5.17}_{-0.74}$ & $0.83^{+0.36}_{-0.36}$ & $58.00^{+0.16}_{-1.17}$ & $2.68^{+0.06}_{-0.47}$ & $14.69^{+1.12}_{-0.10}$ & $2.95^{+0.24}_{-1.20}$ & $3.45^{+0.12}_{-0.43}$ & -- & 386.88/468 & 2332.18 \\
 & Band & -- & $-0.83^{+0.36}_{-0.11}$ & -- & -- & -- & -- & $-3.05^{+0.19}_{-0.30}$ & $30.24^{+2.25}_{-2.33}$ & 371.04/472 & 2338.29 \\
 & CPL & -- & $-0.93^{+0.32}_{-0.15}$ & -- & -- & -- & -- & -- & $32.12^{+2.33}_{-2.53}$ & 373.19/473 & 2322.62 \\
\hline
15.00$\sim$17.00 & FCSYN & -- & -- & -- & -- & -- & $10.01^{+0.96}_{-0.07}$ & $2.62^{+0.35}_{-0.20}$ & -- & 521.06/473 & 3156.38 \\
 & ME\_FCSYN\_CNN & $8.78^{+0.54}_{-2.12}$ & $0.97^{+0.32}_{-0.40}$ & $57.79^{+0.09}_{-1.58}$ & $2.62^{+0.05}_{-0.49}$ & $15.76^{+0.01}_{-1.07}$ & $2.68^{+0.04}_{-1.03}$ & $3.47^{+0.19}_{-0.55}$ & -- & 431.43/468 & 3073.13 \\
 & Band & -- & $-0.95^{+0.36}_{-0.14}$ & -- & -- & -- & -- & $-2.93^{+0.20}_{-0.31}$ & $23.88^{+3.29}_{-2.32}$ & 425.95/472 & 3085.81 \\
 & CPL & -- & $-1.07^{+0.44}_{-0.09}$ & -- & -- & -- & -- & -- & $25.41^{+3.61}_{-1.95}$ & 426.02/473 & 3073.66 \\
\enddata
\tablecomments{For the FCSYN model, $\gamma_{\rm inj}=10^5$ and $\gamma_{\rm cool}=10^2$ are fixed. The magnetic-field parameter of FCSYN is an effective constant-field parameter and is not constrained by the CNN training range of ME\_FCSYN\_CNN.}

\end{deluxetable}
\end{longrotatetable}

The comparison with the standard fast-cooling synchrotron model yields a consistent trend across all time bins. Table~\ref{tab:results} shows that the BIC values of \texttt{ME\_FCSYN\_CNN} are smaller than those of \texttt{FCSYN} in every time bin. This indicates that, even after penalizing model complexity, the fast-cooling synchrotron model in a decaying magnetic field is statistically preferred over the standard fast-cooling synchrotron model with a non-decaying magnetic field. This result suggests that allowing the magnetic field to decay with radius leads to a more self-consistent description of the prompt spectral evolution of this burst.

In terms of computational efficiency, the CNN emulator greatly reduces the fitting cost. On a 16-thread workstation, a full sampling run for a single time bin typically takes only $\sim 400$~s. In contrast, \citet{2016ApJ...816...72Z} reported that fitting a single time bin by directly calling the numerical synchrotron model requires $\sim 160$ CPU hours. Therefore, \texttt{ME\_FCSYN\_CNN} reduces the computational overhead of time-resolved spectral fitting by several orders of magnitude while retaining consistency with the underlying physical model. This makes systematic studies with larger samples and broader parameter-space exploration practically feasible.

\subsection{Comparison with empirical spectral functions}
\label{subsec:empirical_comparison}

We fit the same time-resolved spectra with the Band and CPL functions \citep{1997ApJ...486..928B}. These empirical models are widely used in GRB prompt-emission studies because they flexibly describe the observed spectral curvature, although their parameters are not directly tied to a unique radiation mechanism \citep{2014ApJS..211...12G,2019ApJ...886...20Y,2021ApJ...920...53C}.

\begin{figure}[htbp]
\centering
\includegraphics[width=0.6\textwidth]{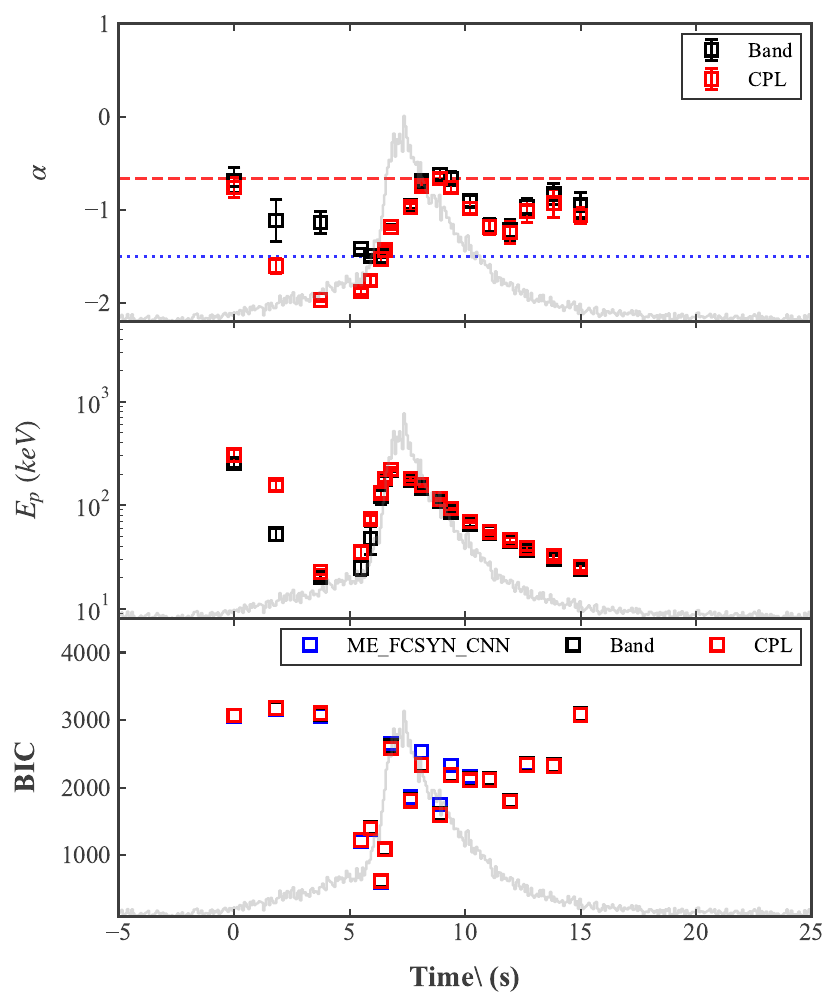}
\caption{Comparison between the empirical Band/CPL functions and the \texttt{ME\_FCSYN\_CNN} model. The top panel shows the temporal evolution of the low-energy photon index $\alpha$ obtained from the Band and CPL fits. The red dashed and blue dotted lines denote $\alpha=-2/3$ and $\alpha=-3/2$, respectively. The middle panel shows the evolution of the spectral peak energy $E_{\rm p}$. The bottom panel shows the BIC values of the \texttt{ME\_FCSYN\_CNN}, Band, and CPL models. The gray curve in each panel shows the prompt-emission light curve for reference.}
\label{fig:empirical_comparison}
\end{figure}

Figure \ref{fig:empirical_comparison} shows the temporal evolution of the empirical spectral parameters and the BIC values of the three models. The Band and CPL fits show clear spectral evolution during the prompt-emission episode. Around the flux peak, the low-energy photon index becomes relatively hard and approaches the synchrotron slow-cooling limit, $\alpha=-2/3$, whereas several off-peak intervals show softer values closer to or below the standard fast-cooling value, $\alpha=-3/2$. The peak energy $E_{\rm p}$ broadly tracks the flux evolution, indicating significant spectral evolution during the main pulse.

The BIC comparison shows that \texttt{ME\_FCSYN\_CNN} is statistically competitive with the empirical Band and CPL functions in many time bins. However, the empirical functions can still provide a better description in some bright intervals. This is not unexpected, because Band and CPL are flexible phenomenological descriptions of spectral curvature, whereas \texttt{ME\_FCSYN\_CNN} is constrained by the electron cooling history in a decaying magnetic field. Similar comparisons have shown that a fast-cooling synchrotron model with magnetic-field decay can reproduce Band-like GRB spectra with a fit quality comparable to that of the empirical Band function \citep{2016ApJ...816...72Z}.

The relatively hard $\alpha$ values around the flux peak also place stronger pressure on a synchrotron-only interpretation. This does not necessarily rule out a decaying-field synchrotron origin, but it suggests that the current synchrotron-plus-adiabatic implementation may be incomplete during the brightest phase. As discussed below, the fitted parameters suggest that the emitting region is generally particle dominated or weakly magnetized in most time bins. In such a regime, additional cooling processes, especially synchrotron self-Compton cooling and Klein--Nishina effects, may become important.

\subsection{Constraints on the prompt-emission jet properties from the fitted parameters}
\label{subsec:jet_properties}

Having verified that \texttt{ME\_FCSYN\_CNN} provides a statistically competitive description of the time-resolved spectra, we next use the fitted physical parameters to constrain the properties of the GRB~231020A jet at the prompt-emission radius. Following previous synchrotron regime studies, we estimate the approximate isotropic-equivalent powers carried by accelerated electrons, magnetic fields, and associated baryons \citep{2008MNRAS.384...33K,2013ApJ...769...69B,2014MNRAS.445.3892B,2015MNRAS.453.1820K}. These quantities are used as local diagnostics of the emitting region, rather than as precise measurements of the initial or global jet composition. In particular, they allow us to examine whether the prompt-emission region is magnetically dominated, moderately magnetized, or particle dominated.

In our model, the electron injection function is written as $Q(\gamma')=Q_0\gamma'^{-p}$, where $\gamma_{\rm inj}<\gamma'<\gamma_{\rm max}$. The comoving injection rate of accelerated electrons is $N'_{\rm inj}=\int_{\gamma_{\rm inj}}^{\gamma_{\rm max}}Q_0\gamma'^{-p}d\gamma'=Q_0(p-1)^{-1}(\gamma_{\rm inj}^{1-p}-\gamma_{\rm max}^{1-p})$, where $p>1$. We estimate the isotropic-equivalent power carried by the accelerated electrons as $L_e\simeq N'_{\rm inj}m_ec^2\gamma'_{inj}\Gamma^2$ \citep{2015MNRAS.453.1820K,2018ApJS..234....3G}. The magnetic power at the emission radius is estimated as $L_B\simeq4\pi R_0^2\Gamma^2c(B_0'^2/8\pi)=R_0^2\Gamma^2cB_0'^2/2$, where $B'_0$ is the comoving magnetic-field strength at $R_0$. Since the baryon loading is not directly constrained by the prompt spectrum, we introduce a proton-to-accelerated-electron number ratio $\xi_p$ and estimate the associated proton kinetic power as $L_p\simeq \xi_p N'_{\rm inj}m_pc^2\Gamma^2$ \citep{2009A&A...498..677B,2011A&A...526A.110D,2013ApJ...769...69B,2018ApJS..234....3G}. The total matter power carried by the radiating electrons and associated cold protons is then defined as $L_k=L_e+L_p$. We define the local effective magnetization parameter of the prompt-emission region as $\sigma_{\rm eff}=L_B/L_k=L_B/(L_e+L_p)$ \citep{2011ApJ...726...90Z}. In addition to $\sigma_{\rm eff}$, we also use $L_e/L_B$ and $L_B/L_p$ as auxiliary diagnostics to show whether the fitted emission region is closer to an electron-dominated, baryon-dominated, or magnetically dominated regime. These ratios should be interpreted as local diagnostics of the prompt-emission region rather than as direct measurements of the initial or global jet magnetization.

\begin{deluxetable*}{ccccccccc}
\tablecaption{Constraints on the prompt-emission jet properties inferred from the fitted \texttt{ME\_FCSYN\_CNN} parameters. \label{tab:jet_properties}}
\tablehead{\colhead{Time bin} & \colhead{$\log N'_{\rm inj}$} & \colhead{$\log L_e$} & \colhead{$\log L_B$} & \colhead{$\log L_p(\xi_p=1)$} & \colhead{$\sigma_{\rm eff}(\xi_p=1)$} & \colhead{$\sigma_{\rm eff}(\xi_p=10)$} & \colhead{$\sigma_{\rm eff}(\xi_p=100)$} & \colhead{$\log(L_e/L_B)$}}
\startdata
0.00--1.82 & 47.43 & 52.49 & 49.06 & 49.91 & 3.67e-04 & 3.58e-04 & 2.91e-04 & 3.43 \\
1.82--3.74 & 48.08 & 51.89 & 47.51 & 50.03 & 4.15e-05 & 3.70e-05 & 1.77e-05 & 4.38 \\
3.74--5.50 & 49.08 & 51.54 & 46.70 & 50.36 & 1.36e-05 & 8.71e-06 & 1.91e-06 & 4.84 \\
5.50--5.91 & 48.70 & 51.45 & 48.26 & 50.21 & 6.19e-04 & 4.13e-04 & 9.53e-05 & 3.18 \\
5.91--6.35 & 49.33 & 51.55 & 51.57 & 50.69 & 0.91 & 0.44 & 0.07 & -0.02 \\
6.35--6.52 & 48.17 & 51.84 & 49.50 & 50.23 & 4.45e-03 & 3.65e-03 & 1.31e-03 & 2.34 \\
6.52--6.78 & 48.14 & 52.04 & 50.66 & 50.65 & 0.04 & 0.03 & 8.27e-03 & 1.38 \\
6.78--7.63 & 48.29 & 52.24 & 48.52 & 50.85 & 1.83e-04 & 1.35e-04 & 3.77e-05 & 3.72 \\
7.63--8.11 & 48.34 & 52.34 & 47.97 & 50.91 & 4.13e-05 & 3.13e-05 & 9.12e-06 & 4.37 \\
8.11--8.90 & 48.31 & 52.34 & 47.16 & 50.80 & 6.41e-06 & 5.12e-06 & 1.71e-06 & 5.18 \\
8.90--9.39 & 47.65 & 52.18 & 47.13 & 50.19 & 8.76e-06 & 8.02e-06 & 4.35e-06 & 5.05 \\
9.39--10.21 & 47.79 & 52.24 & 46.71 & 50.30 & 2.95e-06 & 2.68e-06 & 1.39e-06 & 5.52 \\
10.21--11.05 & 47.56 & 51.77 & 48.70 & 50.14 & 8.43e-04 & 6.99e-04 & 2.59e-04 & 3.06 \\
11.05--11.95 & 47.55 & 51.40 & 49.96 & 50.10 & 0.035 & 0.025 & 6.11e-03 & 1.43 \\
11.95--12.68 & 47.36 & 51.17 & 50.96 & 49.93 & 0.58 & 0.39 & 0.092 & 0.21 \\
12.68--13.83 & 47.98 & 50.91 & 48.52 & 49.62 & 3.86e-03 & 2.68e-03 & 6.62e-04 & 2.39 \\
13.83--15.00 & 47.33 & 50.82 & 50.82 & 49.88 & 0.9 & 0.47 & 0.08 & -0.00 \\
14.50--17.00 & 47.38 & 50.57 & 52.28 & 49.79 & 44 & 19 & 2.9 & -1.71 \\
\enddata
\tablecomments{All powers are isotropic-equivalent and are in units of ${\rm erg~s^{-1}}$. $N'_{\rm inj}$ is in units of ${\rm s^{-1}}$. Logarithms are base 10 for $N'_{\rm inj}$, $L_e$, $L_B$, and $L_p$. The effective magnetization parameter is defined as $\sigma_{\rm eff}=L_B/(L_e+L_p)$. The values of $\sigma_{\rm eff}$ and $L_e/L_B$ are reported in linear form. }
\end{deluxetable*}

The jet physical parameters derived from the spectral fits for each time bin are summarized in Table~\ref{tab:jet_properties}. The inferred ratios indicate that most time bins are not strongly Poynting-flux dominated at the emission radius. Instead, the fitted parameters favor a particle-dominated or weakly magnetized radiating region during most intervals, while several late-time bins show enhanced magnetic importance. This does not contradict the decaying-field synchrotron interpretation, because the key requirement of the model is a decreasing comoving magnetic field during electron cooling, not necessarily $L_B>L_e+L_p$ in every time bin.

A low local magnetic to particle power ratio may indicate that the observed spectra are produced in a post-dissipation or matter-kinetic-dominated stage, where a substantial fraction of the initial magnetic energy has already been converted into particle energy or bulk kinetic energy before or during the prompt-emission episode \citep{2014MNRAS.445.3892B}. Such an interpretation is also qualitatively consistent with magnetic-dissipation models, in which the magnetization can decrease as magnetic energy is released during the prompt-emission phase \citep{2009ApJ...700L..65Z,2011ApJ...726...90Z}.

Moreover, a decaying magnetic field is not exclusive to a strongly Poynting-flux-dominated outflow. In the internal-shock framework, \citet{2006ApJ...653..454P} proposed that the magnetic field generated by shocks may decay on a length scale much shorter than the comoving width of the shocked plasma, thereby alleviating the standard fast-cooling problem of synchrotron radiation. \citet{2014ApJ...780...12Z} further showed that fast-cooling electrons undergoing synchrotron and inverse-Compton cooling in a decaying downstream magnetic field can produce harder low-energy spectra than in the homogeneous-field case. Similarly, \citet{2021Galax...9...68W} found that the magnetic field generated in internal shocks can evolve from an approximately constant phase to a decaying phase, $B'\propto t^{-1}$, which modifies the electron cooling history and hardens the low-energy spectrum. These studies provide a useful interpretation for our low magnetic-to-particle power ratios: even if the emission region is particle dominated, a decreasing magnetic field can still shape the cooling history of fast-cooling electrons and produce harder low-energy spectra. Similar decaying-field effects have also been discussed in large-radius synchrotron models of GRB prompt emission \citep{2014NatPh..10..351U,2018ApJS..234....3G}.

The particle dominated nature inferred for most time bins also points to an important limitation of the present implementation. In such a low-magnetization regime, the photon energy density may become comparable to or larger than the magnetic energy density, and synchrotron self-Compton cooling can affect the electron distribution and the resulting low-energy spectrum. This is particularly relevant because previous numerical studies have shown that different cooling channels, including synchrotron, SSC, and adiabatic cooling, can lead to different low-energy spectral shapes, and that SSC cooling can harden the low-energy spectra under appropriate conditions \citep{2011A&A...526A.110D,2012MNRAS.424.3192B,2014ApJ...780...12Z,2018ApJS..234....3G}. Our current CNN emulator includes synchrotron and adiabatic cooling, but does not self-consistently include SSC cooling or Klein--Nishina effects. We therefore regard this diagnostic result as motivation for extending the emulator to include SSC cooling in future work.

\section{Summary and Discussion}
\label{section5}

This work is motivated by a practical bottleneck in comparing physical radiation models with large observational samples. Although the fast-cooling synchrotron model in a decaying magnetic field provides a physically motivated description of GRB prompt emission, its high numerical cost has limited systematic fitting and statistical testing with \textit{Fermi}/GBM data. To address this issue, we developed a convolutional neural network (CNN)--based spectral emulator for the decaying field fast-cooling synchrotron model. The emulator reduces the wall-clock time for a single spectral calculation from the numerical integration scale to the millisecond level, making Bayesian fitting with physical prompt-emission models computationally feasible.

On the methodological side, we trained the CNN to learn the mapping from the physical parameter vector to the observer-frame photon spectrum. The synthetic training set was generated over a physically motivated parameter space using Latin hypercube sampling, and the model spectra were computed on fixed energy grids appropriate for the NaI and BGO detectors of \textit{Fermi}/GBM. To stabilize the learning problem over broad parameter and flux ranges, we adopted a standardized preprocessing pipeline, including logarithmic transforms for selected physical parameters and a ``logarithmization + energy-bin-wise standardization'' treatment of the output spectra. The detector type was encoded by a one-hot vector, allowing a single network to emulate both NaI and BGO spectral grids.

The independent test-set validation shows that the CNN emulator is nearly unbiased on average, with the mean of the log-spectral residuals close to zero. The typical mean relative error is $\sim 3.88\%$, indicating that the network can reproduce the main spectral features of the numerical model, including slopes, curvature, and peak evolution. We further implemented the trained emulator as a user-defined spectral model in the \texttt{threeML} framework. Since \texttt{threeML} folds the model photon spectrum through the detector response matrix in count space, the CNN emulator can be directly used in standard Bayesian spectral fitting while retaining the computational speed required for time-resolved GRB analysis.

We applied the \texttt{ME\_FCSYN\_CNN} model to the time-resolved spectral analysis of GRB 231020A. The light curve was segmented using Bayesian Blocks, and posterior distributions of the model parameters were obtained through Bayesian inference. The model comparison shows that \texttt{ME\_FCSYN\_CNN} provides systematically better fits than the standard constant-field fast-cooling synchrotron model in terms of BIC and residual behavior. Comparisons with the empirical Band and CPL functions further show that the decaying-field synchrotron model can reproduce the observed spectral shapes with a statistical quality comparable to commonly used empirical functions, while retaining a direct physical interpretation of the fitted parameters. This result supports the use of the decaying-field synchrotron model as a physically interpretable alternative to purely empirical prompt-emission descriptions.

The fitted parameters also provide constraints on the local properties of the emitting region. Using the inferred electron injection rate, bulk Lorentz factor, emission radius, and magnetic field strength, we estimated the electron, magnetic, and proton power components of the jet. These estimates suggest that, although the model is naturally connected to a magnetized dissipation scenario, the local prompt-emission region inferred for GRB~231020A is not strongly magnetically dominated. Instead, the derived energetic ratios are more consistent with a particle-dominated regime in the emission region. This may indicate that a magnetized outflow has already undergone substantial magnetic-energy dissipation by the time the observed prompt radiation is produced. We note, however, that this conclusion depends on the assumed baryon loading and on the simplified radiative treatment. In particular, inverse-Compton cooling and SSC-related effects are not explicitly included in the present calculation and should be considered in future extensions.

The present framework still has several limitations. First, the intrinsic uncertainty of the CNN emulator is not explicitly propagated into the likelihood calculation in the present \texttt{threeML} implementation. The CNN is treated as a deterministic surrogate of the original numerical synchrotron model, while the likelihood accounts for the statistical uncertainties of the observed counts and the background model. To assess the possible impact of this approximation, we performed two additional checks. As the first check, we compared the CNN-emulated spectra with the original numerical spectra at the fitted parameter values of selected GRB~231020A time bins. An example for the high-flux peak interval 6.35--6.52 s is shown in Appendix~\ref{app:emulator_uncertainty}. In this interval, the median and 95th-percentile relative differences between the CNN and numerical spectra are only $0.89\%$ and $1.80\%$, respectively, and the residuals remain at the level of $|\Delta\log_{10}F|\lesssim 0.01$ over most of the energy range. This indicates that the emulator error is small in the high-flux spectral region that dominates the likelihood for this interval. As the second check, we examined the origin of the high-relative-error tail in the test-set validation. The top $5\%$ error points are preferentially located in low-flux spectral-tail regions, where small absolute discrepancies can be amplified into relatively large fractional errors. They are also more common in spectral-tail regions and in some parts of parameter space associated with steeper spectra. Therefore, the quoted 95th-percentile relative error should be interpreted as a conservative global statistic over all valid energy bins and parameter-space regions, rather than as the typical emulator error near the high-flux spectral peak. Nevertheless, because this emulator uncertainty is not explicitly included in the current likelihood, the posterior widths reported in this work should be interpreted as conditional on the trained emulator and may underestimate the full uncertainty budget in some high signal-to-noise intervals. A more complete treatment would propagate the emulator uncertainty into the count-space likelihood, for example through an energy-dependent covariance term, a nuisance systematic component, or an ensemble of independently trained emulators \citep{reiser2025uncertainty}.

Another limitation is that the reliability of the emulator is restricted to the parameter space covered by the synthetic training set. Extrapolation outside this domain is not guaranteed, and future applications to broader GRB samples will require either an expanded training set or an adaptive retraining strategy in poorly sampled regions of parameter space. The current physical model is also confined to a specific decaying-field fast-cooling synchrotron scenario. Several additional physical ingredients may further modify the low-energy spectral index and the inferred energetic partition. For example, under certain conditions, transitions among synchrotron cooling, SSC cooling, and adiabatic cooling can harden the low-energy spectral index \citep{2018ApJS..234....3G}. A more complete treatment of inverse-Compton and SSC cooling may also affect the inferred electron cooling history and the local energy budget. In addition, non-stationary particle injection histories can alter the pile-up behavior of fast-cooling electrons and thereby produce harder low-energy spectra \citep{2021Galax...9...68W,2020ApJ...893L..14L}. Geometric effects, such as variations in viewing angle, Doppler-factor evolution, and equal-arrival-time-surface weighting, may further broaden or soften the observed low-energy spectral-index distribution at the sample level \citep{2023ApJ...947L..11Y,2024ApJ...962...85Y,2025ApJ...986..106G}. Incorporating these effects into the present framework would enable a more comprehensive statistical test of GRB prompt-emission models.

In summary, we have established an efficient computational framework for GRB prompt-emission spectral studies. By enabling millisecond-level spectral emulation of the decaying-field fast-cooling synchrotron numerical model with a CNN, the framework makes it practical to apply a physical synchrotron model in \textit{Fermi}/GBM time-resolved spectral fitting. For GRB~231020A, the CNN emulated decaying-field synchrotron model provides a statistically competitive and physically interpretable description of the observed prompt spectra. The fitted parameters further suggest that the local emitting region is closer to a particle-dominated regime, possibly reflecting substantial magnetic-energy dissipation before or during the prompt-emission phase. This framework lays the computational and methodological foundation for future large-sample tests of magnetic-field decay signatures, distributions of physical jet parameters, and their correlations with empirical GRB observables.

\begin{acknowledgments}
This work is supported by the National Natural Science Foundation of China (NSFC 12233006). Dr. Shan Chang acknowledges support from the National Natural Science Foundation of China 12103046 and the Xingdian Talent Support Plan - Youth Project. We acknowledge the use of the public data from the Fermi data archives. 
\end{acknowledgments}

\vspace{5mm}
\software{Matplotlib \citep{2007CSE.....9...90H}, Numpy \citep{2020Natur.585..357H}, Scipy \citep{2020NatMe..17..261V}, Scikit-learn \citep{scikit-learn}, PyTorch \citep{NEURIPS2019_bdbca288}, 3ML \citep{2015arXiv150708343V}.}

\appendix
\restartappendixnumbering
\section{Emulator uncertainty at fitted parameter points}
\label{app:emulator_uncertainty}

As a supplementary check, we compare the CNN-emulated spectra with the original numerical spectra at the fitted parameter values of selected GRB~231020A time bins. This test is distinct from the independent test-set validation in Section~\ref{section3}, because it directly probes the parameter region used in the observational inference.

\begin{figure}
\centering
\includegraphics[width=0.6\columnwidth]{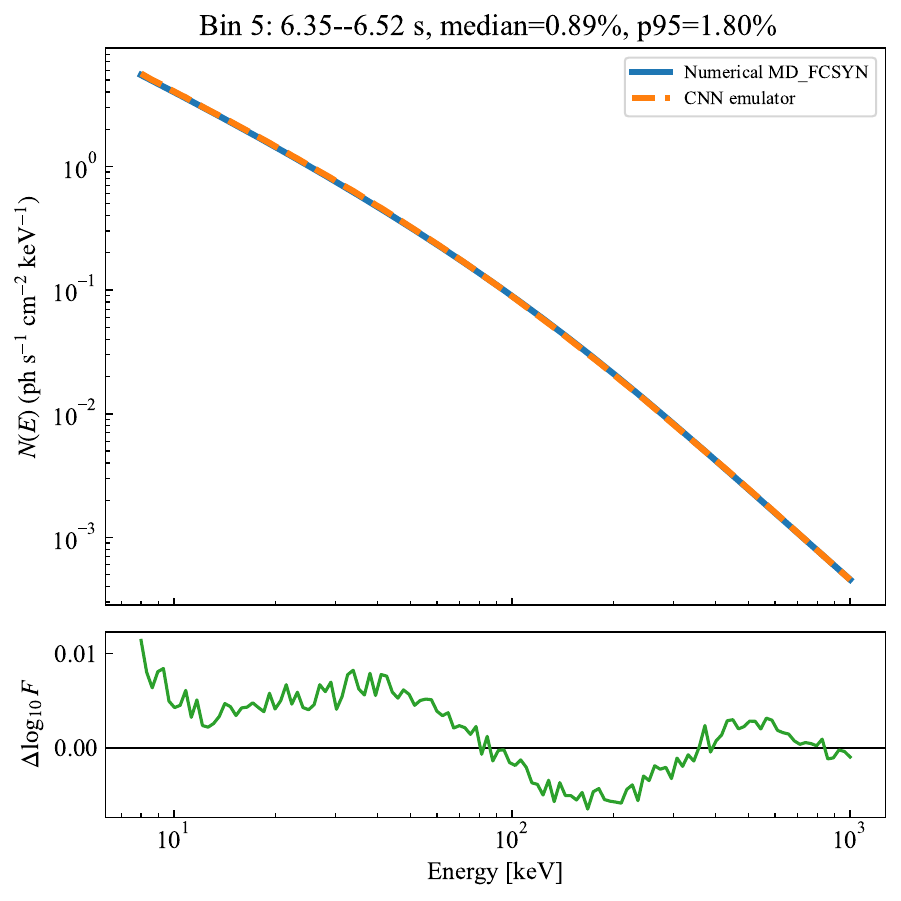}
\caption{
Comparison between the CNN-emulated spectrum and the original numerical spectrum at the fitted parameters of the high-flux peak interval 6.35--6.52 s. The upper panel shows the two spectra, and the lower panel shows the logarithmic residual $\Delta\log_{10}F=\log_{10}F_{\rm CNN}-\log_{10}F_{\rm num}$. The median and 95th-percentile relative differences are $0.89\%$ and $1.80\%$, respectively.
}
\label{fig:cnn_num_peak}
\end{figure}

Figure~\ref{fig:cnn_num_peak} shows the comparison for the high-flux peak interval 6.35--6.52 s. The median and 95th-percentile relative differences between the CNN and numerical spectra are $0.89\%$ and $1.80\%$, respectively. The logarithmic residuals remain at the level of $|\Delta\log_{10}F|\lesssim 0.01$ over most of the energy range. This indicates that the emulator error is small in the high-flux spectral region that dominates the likelihood for this interval.

\bibliography{references}{}
\bibliographystyle{aasjournalv7}
\end{document}